\documentstyle[sprocl]{article}

\input{psfig}

\def\Journal#1#2#3#4{{#1} {\bf #2}, #3 (#4)}

\def\NPB{{\em Nucl. Phys.} B}

\def\PRD{{\em Phys. Rev.} D}

\def\ra{\rightarrow}

\def\be{\begin{equation}}
\def\ee{\end{equation}}
\def\bea{\begin{eqnarray}}
\def\eea{\end{eqnarray}}

\begin{document}

\title{COOLING, MONOPOLES, AND VORTICES IN SU(2) LATTICE GAUGE
THEORY}

\author{JOHN D. STACK, WILLIAM W. TUCKER}
\address{Department of Physics, University of Illinois \\
1110 W. Green St.,Urbana IL 61801, USA\\E-mail: j-stack@uiuc.edu,
wwtucker@students.uiuc.edu} 

\author{ALISTAIR HART}

\address{Department of Physics and Astronomy, University of Edinburgh, \\
Edinburgh EH9 3JZ, Scotland, UK \\E-mail: hart@ph.ed.ac.uk}


\maketitle\abstracts{ We study monopoles and vortices in $SU(2)$ lattice
gauge theory on a $24^4$ lattice at $\beta=2.50$.  We find a value of
fundamental string tension from monopoles in the maximum Abelian gauge
consistent with the full $SU(2)$ value.  Using direct and indirect
center gauges, we find fundamental string tension values from P-vortices which
are larger than the full $SU(2)$ result.  After a single cooling sweep,
the string tensions from monopoles and P-vortices are all 30\% lower 
than the full $SU(2)$ value, while the $U(1)$ string tension in the maximum
Abelian gauge remains consistent with the full $SU(2)$ result.
Blocking the lattice after cooling does not restore the low 
values of string tension found with monopoles and P-vortices.  }

\section{Introduction}
The problem of understanding quark confinement in QCD is as old as
QCD itself-even older, since there was evidence for quarks well before
QCD was precisely formulated.  Among physicists working to understand
confinement, there is universal agreement that the essence of confinement
can be addressed in the pure gauge theory, without light dynamical quarks.
In addition, there is near-universal agreement that the mechanism of
confinement will be `topological' in nature,  caused by a dense gas or
network of topological objects which can disorder Wilson loops and produce
a linear, confining quark potential.

Even without dynamical quarks, there are a host of quantities which 
a theory of confinement must explain.  First and foremost is the
heavy quark potential and in particular, the string tension in the 
fundamental representation.  This has the most real-world relevance
in the spectra of mesons composed of charmed quarks.  
Although only the $SU(3)$ color gauge group is relevant to the real world, the
non-perturbative dynamics of all the $SU(N)$ theories appear to be quite
similar, so as a preliminary to work on $SU(3)$, there has been a 
concerted effort to understand confinement for the simpler case of
an $SU(2)$ gauge group.

The list of topological objects which are possibly relevant to confinement is
short;
instantons, monopoles, and vortices.  Although instantons
have a firm basis as semiclassical objects in the continuum limit, recent
work
casts serious doubt on them as agents of confinement.~\cite{brower}
Accepting this
conclusion leaves 
monopoles and vortices.  
The monopole and vortex approaches to confinement share a common postulate:
namely that the long-range confining physics should be Abelian in character.
They differ in which Abelian subgroup of $SU(N)$ is postulated to carry the
confining physics.  

The path leading to monopoles is normally called `Abelian projection' , in
which
the projection $SU(N) \rightarrow U(1)^{N-1}$ takes place after
gauge-fixing.  Physical quantities may then be calculated using the
projected $U(1)$ fields; this is called `Abelian dominance', or a further
projection made, in which $N-1$ species of magnetic currents of 
monopoles are located and then physical quantities calculated.  This latter
is called `monopole dominance'.

The other topological approach, that leading to vortices, makes the projection
\mbox{$SU(N) \rightarrow Z(N)$,} where $Z(N)$ is the center of $SU(N)$.  
There are
two methods of proceeding.  The one most analogous to Abelian
projection is called `center projection', and uses gauge-fixing followed by
a  projection of
$SU(N)$ links to $Z(N)$ links.~\cite{greensite1,greensite2}  
These $Z(N)$ links are then used to calculate
physical quantities.  Vortices are associated with plaquettes which are
pierced by non-vanishing $Z(N)$ flux.  Here, there is no distinction like
that between Abelian dominance and monopole dominance, since every
$Z(N)$ Wilson loop can be expressed as a product of plaquettes over a surface
which spans the loop.  The second approach to vortices treats a
Wilson loop as a `vortex counter'.
To calculate the heavy quark potential, the Wilson loop is then
simply replaced by its $Z(N)$ part, which is the sign of the trace for $SU(2)$.
This vortex part then carries all the information 
about confinement.~ \cite{tomboulis}  To distinguish these two approaches to
confinement via vortices,
we will refer to vortices located by examining plaquettes after gauge-fixing
and center 
projection as `P-vortices', and simply use the term `vortices' if no 
gauge-fixing or projection at the one-link level is used.  The ideal
situation would be that a P-vortex is  locating the geometrical center
of an actual physical vortex, so in that case results from the two methods
would agree.
 However,
calculations done previously and in this work show that this ideal picture
is too naive.

The present work is devoted to problems with
monopoles and vortices which arise when configurations are smoothed
by cooling.  We restrict ourselves to a discussion of the fundamental
string tension for the case of an $SU(2)$ gauge group.

\clearpage

\section{Gauge-fixing} \label{subsec:g-fix}
The use of gauge-fixing to locate topological objects is common to
both the Abelian and center projection methods.  Since projection 
actually deletes certain dynamical degrees of freedom, the use of a particular
gauge here is different than say an nth order  perturbative calculation of
a gauge invariant quantity. There,
if all terms of a given order are calculated,
one gauge may be more or less convenient than another, but all will give
the same answer in the end.  On the other hand if, as in
Abelian and center projection, certain
dynamical degrees of freedom are removed after gauge-fixing, the resulting
estimate of a physical quantity like the string tension may depend on the
gauge condition used.  Thus even though the quantity being calculated is
gauge-invariant in the full theory, under Abelian or center projection,
there may be an optimal or `maximum' gauge 
which produces the best approximation to the desired physical quantity.
Further, the  gauges commonly used in Abelian and center projection 
involve finding stationary
points of gauge-functionals, and are subject to one form or other of the
Gribov ambiguity.  
The question of how results depend on this ambiguity is important, but 
will not be pursued here.
We use the gauge conditions which have been most successful in previous
calculations, and gauge-fix each configuration once,
i.e. one Gribov copy/configuration
is retained.

\subsection{Maximum Abelian Gauge} \label{subsec:mag}

The maximum Abelian gauge (MAG) will be used for 
Abelian projection in the present
work.  Formulated in the continuum for $SU(2)$, we seek a minimum over
gauge transformations of the
functional

\begin{equation}
G_{mag}=\int[ (A_{\mu}^{1})^{2}+(A_{\mu}^{2})^{2}]d^{4}x,
\label{eq:cont_mag}
\end{equation}
which leads to the following differential condition:
\begin{equation}
(i\partial_{\mu} \pm eA_{\mu}^{3})A_{\mu}^{\pm}=0.
\label{eq:diff_mag}
\end{equation}
The lattice equivalent of minimizing $G_{mag}$ is maximizing the
functional $G_{lmag}$
given by
\begin{equation}
G_{lmag}=\sum_{x,\mu}\frac{tr}{2}\left[U^{\dagger}_{\mu}(x)\sigma_{3}U_{\mu}(x)
\sigma_{3}\right].
\label{eq:lattgf}
\end{equation}
The numerical implementation of the MAG involves a certain stopping criterion.
We used the same criterion as in our previous work.~\cite{stack1}
Expanding the gauge-fixed link
 $U_{\mu}$ in Pauli matrices, 
\begin{equation}
U_{\mu}=U_{\mu}^{0}+i\sum_{k=1}^{3}U_{\mu}^{k}\cdot \sigma_{k},
\end{equation}
we perform the lattice Abelian projection by extracting the U(1) 
link angle $\phi_{\mu}^{3}$ 
\begin{equation}
\phi^{3}_{\mu}=2~\arctan(U_{\mu}^{3}/U_{\mu}^{0}).
\end{equation}
Keeping only the $U(1)$  link formed from 
$\phi_{\mu}^{3}=A_{\mu}^{3}a$ is 
equivalent to retaining only the Abelian field $A_{\mu}^{3}$ in the continuum.
	
The motivation for making the Abelian projection is really monopoles.  
For  a $d=3$ 't Hooft-Polyakov monopole in MAG,
the `charged' fields $A_{\mu}^{1,2}$ are short-ranged, and the Abelian
field $A_{\mu}^{3}$ is long-ranged and resembles that of a Dirac monopole
with two Dirac units of charge, $eg=4\pi$.  The Dirac string that appears in
this gauge 
is the basis of the Toussaint-DeGrand method
of monopole location on the lattice.~\cite{toussaint}  
In $d=3$, a monopole is at the end of
a Dirac string, while in $d=4$ the magnetic current of a monopole lies on
the edge of a Dirac sheet.

\subsection{Direct and Indirect Center Gauges} \label{subsec:centg}

In what is called the direct center Gauge (DCG), the following functional is
maximized over gauge transformations:
\be
G_{dcg}= \sum_{x,\mu} (tr(U_{\mu}(x)))^{2}.
\label{eq:dcg}
\ee
Using the relation between the trace of group element
matrices in fundamental and adjoint 
representations,
\be
tr(U_{A})=(tr(U_{F}))^{2}-1,
\ee
we see that DCG condition maximizes the trace in the adjoint
representation. For small gauge fields this is the same as minimizing
\be
\sum_{x,\mu} (A_{\mu}^{a})^{2},
\ee
which is the Landau gauge condition.  
The functional $G_{dcg}$ in Eq.(\ref{eq:dcg}) is `center-blind' , i.e.
invariant to $U_{\mu}\ra Z\cdot U_{\mu}$, where $Z$ is a member of the
center group, so the DCG can be
thought of as a center-blind Landau gauge.  All components of $A_{\mu}^{a}$
are suppressed as much as possible, 
modulo a center factor in the fundamental representation
links.

A variant on DCG is the indirect center Gauge (ICG), where after a preliminary
gauge-fixing to MAG, the functional
\be
G_{icg}=\sum_{x,\mu} (\cos(\phi_{\mu}^{3}))^{2}
\ee
is maximized over $U(1)$ gauge transformations.  Having suppressed 
$A_{\mu}^{1,2}$ by the use of MAG, this is a center-blind way to 
finally suppress $A_{\mu}^{3}$.

For both DCG and ICG, the center projection is  done by writing
\be
U_{\mu}(x)=sign(tr(U_{\mu}(x))) \cdot \bar{U}_{\mu}(x).  
\ee
Non-perturbative, confining physics is postulated to reside in the
$Z(2)$ gauge field $Z_{\mu}=sign(tr(U_{\mu}))$. The
presence of a P-vortex is signified by a negative $Z(2)$ plaquette
which is supposed to represent the physical center of an actual vortex.
In our numerical calculations, the stopping criteria used for DCG and ICG
were of a similar nature and quality to that used for MAG.

\section{Unsmoothed Results } \label{subsec:unsmooth}

\begin{figure}[t]
\begin{center}
\psfig{figure=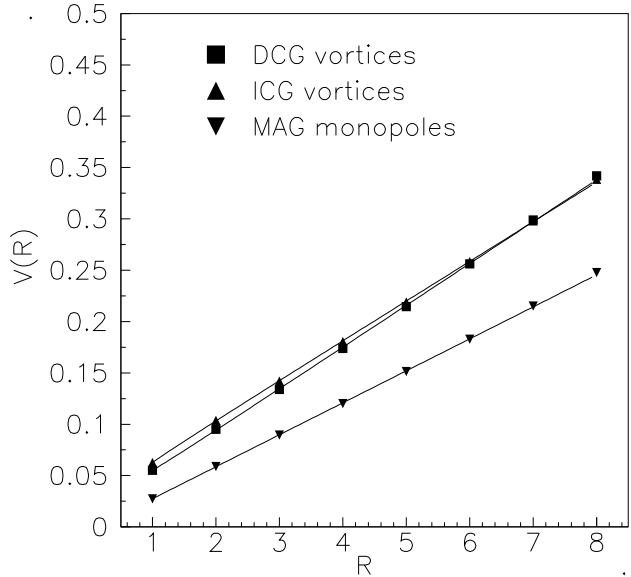,height=3.0in}
\end{center}
\caption{Heavy quark potentials from monopoles (MAG),
and vortices (DCG and ICG).  \label{fig:uncool_unblock}}
\end{figure}

The calculations presented here are for the Wilson form of
$SU(2)$ lattice gauge theory, at $\beta=2.50$, on a $24^4$ lattice.
We have analyzed $49$ configurations in  the MAG, and $30$ in DCG and ICG.
In these configurations, we have extracted heavy quark potentials, and
examined some features of the distribution of monopoles and P-vortices.
As mentioned, we take one Gribov copy/configuration. 
The number of configurations is moderate, but
as seen in Fig.(\ref{fig:uncool_unblock}), the potentials are extremely 
linear and noise-free,
a characteristic of calculations with topological objects.  The string
tension in the fundamental representation is easily extracted from the
slopes of the potentials vs R.  The results are tabulated in 
Table~\ref{tab:uncool}.
The corresponding full $SU(2)$ fundamental string tension at $\beta=2.50$ 
is 0.033(2) from our own previous work~\cite{stack1} on $16^4$, or 
0.0325(12) from Bali, {\it et al}~\cite{bali1} on $ 32^4$, at 
$\beta=2.5115$.  The figures in Table~\ref{tab:uncool}
show that the monopole string tension
is very consistent with the full $SU(2)$ results, but that the 
P-vortex string tensions are too large, by an amount outside error bars.
A high value of $\sigma_{dcg}$ and $\sigma_{icg}$ for one Gribov 
copy/configuration was also recently observed by Bornyakov {\it et al }
.~\cite{bornyakov}

\begin{table}[t]
\caption{$\beta=2.5$, $24^4$ Wilson Action String Tensions}
\label{tab:uncool}
\vspace{0.2cm}
\begin{center}
\footnotesize
\begin{tabular}{|c|c|c|}
\hline
$\sigma_{mag}(mon)$  & $\sigma_{dcg}$   & $\sigma_{icg}$\\
\hline
0.031(1)  &  0.040(1)  &  0.039(1) \\
\hline
\end{tabular}
\end{center}
\end{table}


Turning to the distribution of monopoles and P-vortices, the percentage
of links with magnetic current is 1.36(1), which means there are 
$\sim 18,000$ links with magnetic current in a typical configuration.
The largest cluster has an average size of 7554(124) links.  Only the latter
is relevant to confinement so really only $\sim 0.57\%$ of the links on
the lattice play a role in 
the confining part of the magnetic current.

For P-vortices, the percentage of links pierced by a P-vortex is
3.21(1) for DCG and 3.78(1) for ICG.  In the indirect center gauge, there
is a strong correlation between the locations of magnetic current and 
of P-vortices.  The magnetic current $m_{\mu}$ resides on the dual of the
original lattice, so the timelike magnetic current $m_{t}$
may be placed at the center of a spacial $(xyz)$ cube of the original lattice,
$m_{x}$ at the center of a $yzt$ cube, etc for $m_{y},m_{z}$.
As first noted by Greensite {\it et al}~\cite{greensite}, a large 
percentage of the time when
a 3-cube contains magnetic current, 
two of its faces are pierced by P-vortices.
We measured this percentage
and find 93(1)\% for this result, 
consistent with Greensite {\it et al}.~\cite{greensite} 
In other words, in the ICG, P-vortices end on
monopoles.  Since P-vortices {\it are} visible, i.e. they cause minus signs
in fundamental Wilson loops, they resemble Dirac strings
(or sheets in $d=4$) with $2\pi$ rather than $4\pi$ units of flux.  
A fundamental quark in a Wilson loop acts like a half-integer charge, so
a vortex threading the loop gives a factor $\exp(i2\pi/2)=-1$.
Now
`clock' or Z(n) approximations to $U(1)$ are often successfully used 
with rather large
values of $n$.
The ICG is an extreme form of this
approximation
where $U(1)$ is projected to $Z(2)$.   The Coulomb flux of
a monopole gets squeezed into  a $Z(2)$ vortex passing through the monopole.
This shows up on the lattice as two negative $Z(2)$ plaquettes on the
cube-faces surrounding the monopole.

\section{Smoothed Results } \label{sec:smooth}

Not all the information contained in a sequence of configurations
generated in a Wilson action simulation is relevant to the physics of
confinement.  Local smoothing  of configurations is a way to suppress
ultraviolet fluctuations, while keeping the long-range physics intact.
In the present work,  our smoothing operation is a single cooling sweep
of the lattice, in which each link is replaced by its action-minimizing
value in the fixed environment of nearby links or `staples'.  

The string tension is stable under a single cooling, showing that the
cooled configurations still encode the information about confinement 
present in the original configurations.  It is reasonable to demand that 
a description of the confining degrees of freedom also be stable under
cooling.  To test this we subjected the once-cooled configurations to
the same gauge-fixing and object location procedures used in the
previous section.  The results are shown in Table~\ref{tab:cool}.  

\begin{table}[t]
\caption{Cooled Wilson action string tensions}
\label{tab:cool}
\vspace{0.2cm}
\begin{center}
\footnotesize
\begin{tabular}{|c|c|c|}
\hline
$\sigma_{mag}(mon)$  & $\sigma_{dcg}$   & $\sigma_{icg}$\\
\hline
0.021(1)  &  0.021(1)  &  0.022(1) \\
\hline
\end{tabular}
\end{center}
\end{table}

\begin{figure}[t]
\begin{center}
\psfig{figure=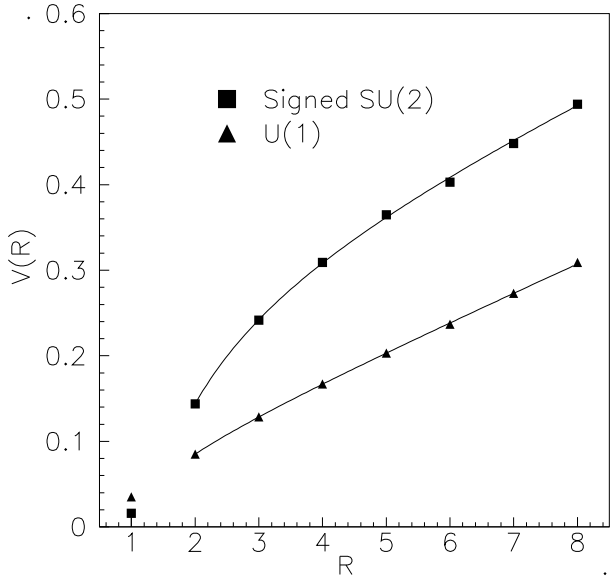,height=3.0in}
\end{center}
\caption{Potentials after cooling from $U(1)$ (MAG) and sign of Wilson loop 
\label{fig:u1_sign}}
\end{figure}
The string tensions
from monopoles, and vortices from DCG and ICG now agree,  but all are 
$\sim 30\%$ low, compared to the full $SU(2)$ result.  
This poses a 
serious problem for claims that monopoles found after gauge-fixing to the
MAG, or P-vortices found after gauge-fixing to either DCG or ICG  are a
correct identification of the infrared confining degrees of freedom.

It is of interest to see how the numbers and distribution of monopoles
and P-vortices are affected by cooling.  In DCG there
are now only $1.7\%$ of plaquettes pierced by a P-vortex, an almost $50\%$
reduction compared to uncooled configurations.  There is a similar reduction of
the number of P-vortices in ICG.  

For monopoles deduced from the MAG, 
the reduction  in number
after cooling is even more dramatic-now only $0.16\%$ of the links carry
magnetic current, a reduction by a factor of $\sim 8.4$.  Whereas before
cooling the largest cluster of magnetic current contained $\sim 7500$ links,
after cooling the largest cluster contains only $\sim 1200$ links.
While $30\%$ of the string tension has been lost, from another viewpoint,
it is remarkable that $70\%$ of the string tension can be obtained with
a magnetic current occupying only $0.16\%$ of the links of the lattice.
Cooling also heavily suppresses small clusters, which are known to be
irrelevant for confinement.  The connection between P-vortices in ICG and
MAG monopoles is even tighter; now 98(1)\% of the time a monopole cube-face
is pierced by two P-vortices.

The results just discussed show that the most straightforward 
application of gauge-fixing and object location methods are not stable
under smoothing.  It does not follow that all hope of a topological
description of confinement is lost.  From the viewpoint of vortices, we
may use the second method of proceeding with vortices.  This uses no 
gauge-fixing, instead the potential is calculated by replacing the Wilson
loop by its sign,
loop by loop.  This yields a fundamental
string tension of 0.031(1),  
consistent with the full $SU(2)$ result.  From the viewpoint of Abelian
projection, we may gauge-fix to MAG,  and calculate the potential from the
$U(1)$ links directly, without a further reduction to monopoles.  This yields
a string tension after cooling of 0.034(1),  slightly high, but still
consistent with the
full $SU(2)$ numbers.  The potentials from these two calculations
are shown in Fig.~(\ref{fig:u1_sign}).   (The differing short range or
Coulombic terms in the two potentials is easily explained by the fact
that perturbative exchange of charged gluons is suppressed in the MAG.)

Putting these last two results together suggests that the problem lies
neither with the idea of a topological explanation of confinement nor
the use of gauge-fixing, but rather with our methods of location of
topological objects.

\section{Extended Objects}
As just discussed, the $U(1)$ field obtained via Abelian projection on
cooled lattices retains the full $SU(2)$ string tension.  The usual 
expectation is that the confining part of such a $U(1)$ field can be
characterized in terms of the magnetic current of monopoles.  However,
our attempts to find this current have failed so far.  In addition to
the standard approach described in Sec.~\ref{sec:smooth}, we have made
two other attempts, both looking for monopoles on a larger scale.  The
first method begins by casting the cooled $SU(2)$ configurations into
the MAG as before, but then looking for `extended' monopoles by applying
the Toussaint-DeGrand method to 2-cubes, rather than 1-cubes.~ \cite{stack2}  
A second
attempt was to block the cooled $SU(2)$ configurations in 
a standard way~\cite{stack2},
gauge-fix to 
the MAG on the blocked lattice, followed by monopole location, also on the
blocked lattice.  In both attempts, the resulting monopole string tension is
still $\sim 30\%$ lower than the corresponding full $SU(2)$ string tension.
(Our previous work~\cite{stack2}  
on a $20^4$ lattice held out some hope that blocking
would restore the lost string tension, but the present work on $24^4$ does
not support this.)

Likewise, although the signed Wilson loops carry the full $SU(2)$ string 
tension, we have been unable to characterize these signed loops in terms
of P-vortices.  We have also applied blocking here, taking the cooled,
blocked $SU(2)$ configurations into DCG and ICG.  The resulting P-vortex
string tensions are again $\sim 30\%$ low.

To summarize, on the positive side, we have shown that 
the maximum Abelian gauge and Abelian projection itself survive cooling, as
does the method of dealing with vortices which does not try to pin down
their locations.  However, we have also shown,
on a suitably large lattice, that present methods
of locating topological objects are unstable against smoothing.  This is a
serious problem for claims that these objects correctly identify the
confining degrees of freedom
in $SU(2)$ lattice gauge theory.


\section*{Acknowledgments}
The work of J. Stack and W. Tucker was supported by the U.S. National
Science Foundation.  The work of A. Hart was supported by PPARC (UK).

\end{document}